\documentclass[english,10pt]{article}

\usepackage[left=.86in,right=.86in,bottom=.86in,top=.86in]{geometry}

\usepackage[utf8]{inputenc}
\usepackage{babel}
\usepackage{slantsc}
\usepackage{array}
\usepackage{amsmath}
\usepackage{amsthm}
\usepackage{amssymb}
\usepackage{graphicx}
\usepackage{enumerate}
\usepackage{mathtools}
\usepackage{natbib}
\usepackage{bbm}
\usepackage{tikz}
\usepackage{multirow}
\usetikzlibrary{tikzmark}
\usepackage{subfigure}
\usepackage{xspace}
\usepackage{bigints}
\usepackage{booktabs}
\usepackage{thmtools}

\usepackage{pgf}
\usepackage{pgfplots}
\pgfplotsset{compat=newest} 
\pgfplotsset{plot coordinates/math parser=false} 
\newlength\figureheight 
\newlength\figurewidth 

\usepackage{heuristica}
\usepackage[heuristica,vvarbb,bigdelims]{newtxmath}
\usepackage[T1]{fontenc}

\newtheorem{theorem}{Theorem}

\newtheorem{corollary}{Corollary}

\newtheorem{lemma}{Lemma}

\newtheorem*{theorem*}{Theorem}
\theoremstyle{definition}
\newtheorem{definition}{Definition}
\declaretheorem[style=definition]{example}
\renewcommand\thmcontinues[1]{continued}

\usepackage[labelsep=period]{caption}

\DeclareMathOperator*{\supp}{supp}

\makeatletter
\newcommand{\subalign}[1]{%
  \vcenter{%
    \Let@ \restore@math@cr \default@tag
    \baselineskip\fontdimen10 \scriptfont\tw@
    \advance\baselineskip\fontdimen12 \scriptfont\tw@
    \lineskip\thr@@\fontdimen8 \scriptfont\thr@@
    \lineskiplimit\lineskip
    \ialign{\hfil$\m@th\scriptstyle##$&$\m@th\scriptstyle{}##$\crcr
      #1\crcr
    }%
  }
}
\makeatother

\definecolor{PennBlue}{RGB}{001,031,091}
\definecolor{PennRed}{RGB}{153,0,0}
\usepackage{hyperref}
\hypersetup{
	pdfborder = {0 0 0},
    colorlinks,
    citecolor=PennRed,
    filecolor=PennRed,
    linkcolor=PennRed,
    urlcolor=PennRed,
    breaklinks=true
}

\usepackage[blocks]{authblk}
\title{\fontsize{20}{20} 
\selectfont{The Empirical Content of Bayesianism}\thanks{\fontsize{10}{10}\selectfont\baselineskip0.42cm This paper is based on the second chapter of my PhD dissertation at MIT. An earlier version of this paper was previously circulated under the title ``Tests of Bayesian Rationality.'' I am grateful to Daron Acemoglu, Marios Angeletos, Sandeep Baliga, Drew Fudenberg, Arda Gitmez, Parag Pathak, Larry Samuelson, Alvaro Sandroni, Eran Shmaya, Alireza Tahbaz-Salehi, Juuso Toikka, Olivier Wang, Iv\'{a}n Werning, Alex Wolitzky, Muhammet Yildiz, and seminar participants at MIT, the Becker Friedman Institute, and Aalto University for their valuable comments.}
}
\author{\vspace{0.1in}\hspace{0.0in}\fontsize{16}{16}\selectfont  {Pooya Molavi}\thanks{\fontsize{10}{10}\selectfont Northwestern University, \href{mailto:pmolavi@kellogg.northwestern.edu}{pmolavi@kellogg.northwestern.edu}.}
}
\date{\today
}

\AtBeginDocument{%
  \mathchardef\mathcomma\mathcode`\,
  \mathcode`\,="8000 
}
{\catcode`,=\active
  \gdef,{\mathcomma\discretionary{}{}{}}
}

\vfuzz \hfuzz
\allowdisplaybreaks

\newcolumntype{C}[1]{>{\centering\arraybackslash$}p{#1}<{$}}
\newlength{\mycolwd}

\begin{document}

\maketitle

\begin{abstract}
\fontsize{11}{11}\selectfont \baselineskip0.58cm
\noindent This paper characterizes the conditions under which the observed beliefs of a group of agents are consistent with Bayesian updating. Beliefs are consistent with Bayesianism if they arise from the application of Bayes' rule given some subjective distribution for the state and the signals agents observe between periods. The paper's main finding is that beliefs are consistent with Bayesianism if and only if the mean of the distribution of posteriors is uniformly absolutely continuous with respect to the prior. Furthermore, the paper shows that the existing results on the empirical content of Bayesianism rely on additional restrictions on permissible subjective distributions, such as the requirement that agents have correct beliefs about the distribution of signals.
\end{abstract}

\thispagestyle{empty}
\newpage 
\setcounter{page}{1}\fontsize{12}{12}\selectfont\baselineskip0.695cm

\section{Introduction}
Following the treatise of \citet*{savage1972foundations}, the Bayesian theory of probability has become the dominant paradigm in the modeling of decision-making under uncertainty. This paradigm's dominance in economics is not unwarranted. It allows one to assign probabilities to unique or rare events. It has an elegant foundation in the study of rational choice under uncertainty. And it is appealing from a normative perspective---as \citet*{epstein1993dynamically} proclaim, ``dynamically consistent beliefs must be Bayesian.'' What is less clear is whether Bayesianism is a good positive model of individual behavior. Settling this question requires characterizing the testable predictions of Bayesian rationality.

This paper characterizes the empirical content of Bayesianism. It considers an analyst who observes how the beliefs of a group of agents evolve. The analyst can perfectly observe agents' beliefs about an arbitrary state of the world but not how agents plan to update their beliefs based on their signals. The analyst aims to determine if the observed belief sequence is consistent with Bayesian updating given a joint subjective distribution for the state and signal. The paper's main result is that the analyst can rationalize his observation as consistent with Bayesianism if and only if the mean of the distribution of posteriors is uniformly absolutely continuous with respect to the prior.

This result suggests that Bayesianism has limited testable implications. In particular, when the state space is finite, a belief sequence is consistent with Bayesian updating if and only if the posterior mean is supported on a subset of the prior's support. If the prior has full support on a finite state space, then Bayesianism imposes \emph{no} restrictions on the distribution of posteriors. These results question the feasibility of testing Bayesianism in observational data.

The paper's characterization result is obtained under assumptions that make it \emph{easier} for the analyst to reject agents' Bayesianism. The analyst is assumed to directly observe agents' beliefs, instead of having to infer them from their actions. The analyst is free to elicit what agents believe about an arbitrary state belonging to an arbitrary state space, and he can observe those beliefs without any observation noise. I further assume that there is a large number of ex ante identical agents, and agents observe i.i.d. signals. All these assumptions make it easier to disprove agents' Bayesianism; yet, any observation meeting the uniform absolute continuity condition can be rationalized. However, the assumptions make it possible to show that this condition is both necessary and sufficient for consistency with Bayesianism.

This finding may appear at odds with the existing results in the literature. \cite*{aumann1995repeated} and \cite*{Kamenica2009} argue that a belief sequence is Bayes' plausible if and only if the posterior mean equals the prior. \citet*{Shmaya2016} argue that any belief sequence in which the prior is in the relative interior of the convex hull of posteriors is consistent with agents' use of Bayes' rule. Both of these conditions are more restrictive than the uniform absolute continuity condition derived in this paper. However, those results are obtained under additional restrictions on what constitutes a reasonable subjective belief for agents.

I prove two additional theorems to clarify the relationship between this paper's characterization and those in the literature. The theorems adapt the existing results to the more general setting of the current paper, thus making them directly comparable to the paper's result. They demonstrate that the earlier results characterize the empirical content of Bayesianism only under additional assumptions on agents' subjective beliefs. \cite*{aumann1995repeated} and \cite*{Kamenica2009} do so by requiring agents to have correct beliefs about the distribution of signals, whereas \cite*{Shmaya2016} require the subjective belief to have the same support as the true distribution.

Beyond the literature discussed above, this paper also contributes to the literature on deviations from rational expectations. This literature can be roughly divided into two strands. The first strand, such as \citet*{esponda2016berk, esponda2021equilibrium}, maintains the assumption of Bayesian updating but admits the possibility that agents hold misspecified priors.\footnote{See also \citet*{bohren2016informational}, \citet*{fudenberg2017active}, \citet*{frick2020misinterpreting}, \citet*{fudenberg2021limit}, and the references therein.} The second strand studies the implications of non-Bayesian updating rules such as representativeness and availability heuristics \citep*{tversky1974judgment}, confirmation bias \citep*{rabin1999first}, and diagnostic expectations \citep*{bordalo2018diagnostic}.\footnote{See \citet*{epstein2010non}, \citet*{molavi2018theory}, and \citet*{cripps2018divisible} for other examples of non-Bayesian updating rules.} This paper's main result clarifies the relationship between these two strands of the literature by showing that almost any non-Bayesian updating rule is observationally equivalent to Bayesian updating given a misspecified prior about the distribution of signals. \citet*{Hauster_Bohren_2021} also study the question of when non-Bayesian updating rules can be represented as misspecification. They extend \citet*{Shmaya2016}'s analysis by making agents' forecasts of their future beliefs observable and deriving necessary and sufficient conditions for an updating rule and a forecast to have a misspecified-model representation. In contrast, this paper's focus is characterizing the empirical content of Bayesianism absent information on agents' beliefs about how they will update their beliefs.

\section{Setup}\label{sec:setup}

This section introduces the environment and defines what it means for belief sequences to be consistent with Bayesianism.

\subsection{The Environment}
I consider an analyst who examines whether a group of agents updates their beliefs about a fixed state of the world using Bayes' rule. The state is denoted by $x$ and belongs to a separable metric space $X$.

There is a large number of agents indexed by $i\in I$. In each of two periods $t=0,1$, the analyst elicits what each agent believes about the state of the world. Agent $i$'s time-$t$ belief about the value of $x$ is a probability distribution, denoted by $\mu_{it}\in\Delta(X)$. I assume that the analyst can perfectly observe $\mu_{it}$ for all $i$ and $t=0,1$. 

Agents' beliefs might evolve between the periods due to new information. I let $s_i$ denote the signal observed by agent $i$ between the two periods and assume without loss of generality that agents' signals belong to the set $S\equiv \Delta (X)$. Each agent $i$ uses a measurable mapping $\varphi_i:s_i\mapsto \mu_{i1}$ to form her posterior based on her realized signal $s_i$.\footnote{The term ``posterior'' here refers to agents' beliefs at time one, regardless of whether those beliefs are derived from agents' prior via Bayes' rule.} \footnote{The assumption that an agent's posterior is a \emph{deterministic} function of her signal is without loss of generality. Any random updating rule is equivalent to a deterministic updating rule with a true distribution  of signals $\mathbb{P}$ that does the randomization for agents.}

I make several assumptions that all help the analyst conclude that agents must \emph{not} be Bayesian---these assumptions make the negative result of the paper even more striking. First, agents are ex ante identical. In particular, $\mu_{i0}=\mu_0^*$ for some $\mu_0^*\in\Delta(X)$ and all $i\in I$. Second, agents' signals are independent and identically distributed, with $\mathbb{P}\in\Delta(S)$ denoting the true distribution of signals given the fixed state of the world. Third, agents all use the same mapping $\varphi=\varphi_i$ to form their beliefs as a function of their signals. Fourth, the number of agents is large enough that the empirical distribution of observed posteriors $\{\mu_{i1}\}_{i\in I}$ provides an arbitrarily good approximation to the corresponding population distribution. 
Specifically, I assume that the analyst can perfectly observe the population distribution of agents' posterior beliefs, denoted by $F_1^*\in\Delta(\Delta (X))$. Finally, the analyst is assumed to know everything described so far. While these assumptions lead to a tight characterization result, they are not required for the paper's finding that Bayesianism only imposes a weak restriction on belief sequences. Section \ref{sec:assumptions_discussion} elaborates on this point.

The analyst's question is whether he can interpret the pair $(\mu_0^*,F_1^*)$ as being consistent with Bayesian updating given some subjective belief held by agents. The paper's main result establishes that any pair $(\mu_0^*,F_1^*)$ that satisfies an absolute continuity condition can be rationalized as consistent with Bayes' rule. This result is obtained despite the aforementioned assumptions being biased towards rejecting Bayesianism.

\subsection{Bayes Plausibility}
Before presenting the main result, it is necessary to define what it means for an observed pair $(\mu_0^*,F_1^*)$ to be consistent with Bayesianism. Agents are Bayesian if they (i) possess a well-defined subjective belief over the set of state-signal pairs; (ii) assign positive probabilities to signals that occur with positive probabilities; and (iii) update their beliefs following each signal that occurs with a positive probability using Bayes' rule.

The first two requirements can be formalized in the following way: First, agents must have a subjective distribution $\mathbb{Q}\in\Delta(X\times S)$ over the set of states and signals.\footnote{Note that the subjective distribution $\mathbb{Q}$ does not have an $i$ subscript since I have assumed that agents are ex ante identical.} Second, their subjective distribution must assign a positive probability to any signal that is realized with a positive probability. This is to ensure that agents can use Bayes' rule following every contingency. This requirement can be expressed using the following notion:
\begin{definition}
If $P$ and $Q$ are probability distributions over the same measurable space, $P$ is \emph{uniformly absolutely continuous} with respect to $Q$ if there exists a positive constant $c$ such that $P(E)\leq cQ(E)$ for any measurable set $E$.\footnote{See Lemma \ref{lem:unif_abs_cont} of the appendix for an equivalent way of defining uniform absolute continuity.}
\end{definition}

I require the subjective distribution $\mathbb{Q}$ to be such that the true distribution of signals $\mathbb{P}$ is uniformly absolutely continuous with respect to the $S$-marginal $\mathbb{Q}_S$. When $\mathbb{P}$ has a finite support, this condition reduces to the requirement that $\mathbb{Q}$ does not rule out any signal that is realized with a positive probability. More generally, the uniform absolute continuity requirement ensures that agents' beliefs do not disproportionately discount the likelihood of certain signals. Although this condition is not necessary for the paper's main finding, having a more demanding notion of Bayesianism strengthens the result by highlighting the fact that the conclusion does not rely on the inapplicability of Bayes' rule after zero-probability events. It also allows me to turn the statement of the main result into an ``if and only if'' statement.

The third criterion for Bayesianism is agents' use of Bayes' rule to update their beliefs. This criterion is formally expressed through the concept of regular conditional probability. Given the measurable space $(X\times S,\mathcal{X}\times\mathcal{S})$ and probability distribution $\mathbb{Q}\in\Delta(X\times S)$, a \emph{regular conditional probability} of $\mathbb{Q}$ given $\mathcal{S}$ is a mapping $\nu:S\times\mathcal{X}\to[0,1]$ such that (i) $\nu(s,\cdot)$ is a probability distribution on $X$ for every $s\in S$, (ii) the mapping $s\mapsto\nu(s,D)$ is measurable for all $D\in\mathcal{X}$, and (iii) the kernel $\nu$ satisfies
\begin{equation}\label{eq:Bayes_rule}
\mathbb{Q}(D\times E) = \int_E \nu(s,D)\mathbb{Q}_S(ds)
\end{equation}
for all $D\in\mathcal{X}$ and $E\in \mathcal{S}$, where $\mathbb{Q}_S$ is the $S$-marginal of $\mathbb{Q}$. The regular conditional probability $\nu$ defines a mapping $\varphi:s\mapsto \nu(s,\cdot)$ from agents' signals to their posteriors. Agents are Bayesian given subjective distribution $\mathbb{Q}$ if they use this mapping to update their beliefs.

The regular conditional probability $\nu$ determines agents' posterior beliefs as a function of their subjective belief and the realized signal. However, it does not specify the distribution of those posterior beliefs. In particular, for any event $D\in\mathcal{X}$, $\nu(s,D)$ is a random variable whose distribution depends on the distribution of the signal $s$. To determine the probability with which each posterior is realized, one needs to use the true distribution of signals. Given a regular conditional probability $\nu$ and the true distribution of signals $\mathbb{P}$, agents' posterior about the state $x$ is distributed according to the probability distribution $F_\nu\in\Delta(\Delta (X))$, defined as
\begin{equation}\label{eq:population_dist}
F_\nu(\{\mu_1\in \Delta(X):\mu_1\in E\}) \equiv \mathbb{P}\left(\{s\in S:\nu(s,\cdot)\in E\}\right)
\end{equation}
for all $E\in\mathcal{S}$. This is the observed distribution of posteriors if agents are Bayesian with subjective distribution $\mathbb{Q}$ (and the corresponding $\nu$) and the true distribution of signals is given by $\mathbb{P}$. 

I can now define what it means for agents' observed belief sequence to be consistent with Bayesianism.
\begin{definition}\label{def:Bayesianism}
Given the state space $X$, signal space $S=\Delta(X)$, and true distribution of signals $\mathbb{P}\in\Delta(S)$, a pair of observations $(\mu_0^*,F_1^*)$, consisting of agents' prior and the distribution of their posteriors about the state, is \emph{consistent with Bayesianism} if there exists a subjective distribution $\mathbb{Q}\in\Delta(X\times S)$ for agents that satisfies the following conditions:
\begin{itemize}
    \item[(a)] $\mathbb{Q}_X=\mu_0^*$,
    \item[(b)] $\mathbb{P}$ is uniformly absolutely continuous with respect to $\mathbb{Q}_S$,
    \item[(c)] $\mathbb{Q}$ has a regular conditional probability $\nu$ such that $F_\nu=F_1^*$,    
\end{itemize}
where $\mathbb{Q}_X$ and $\mathbb{Q}_S$ are the $X$- and $S$-marginals of the subjective distribution $\mathbb{Q}$, respectively, and $F_\nu$ is the distribution of posteriors defined in \eqref{eq:population_dist}.
\end{definition}

This definition formalizes the intuitive notion of Bayesianism laid out at the beginning of this subsection. The analyst's task is to find a subjective distribution $\mathbb{Q}$ that explains the observed changes in agents' beliefs. This conjectured $\mathbb{Q}$ is a joint distribution for the state and signal that must satisfy three conditions: Condition (a) of Definition \ref{def:Bayesianism} simply requires the conjectured distribution to be consistent with the observed prior. Condition (b) is the requirement that agents assign non-vanishing probabilities to signals that are realized with positive probabilities. Condition (c) requires that the observed distribution of posteriors matches the distribution obtained when agents start with the conjectured distribution $\mathbb{Q}$, observe signals as per $\mathbb{P}$, and update their beliefs using Bayes' rule.

\section{The Empirical Content of Bayesianism}\label{sec:result}
\subsection{The Main Result}
The paper's main result provides a necessary and sufficient condition for the observed pair $(\mu_0^*,F_1^*)$ to be consistent with Bayesianism: 
\begin{theorem}\label{thm:main}
A pair $(\mu_0^*,F_1^*)$ is consistent with Bayesianism if and only if the mean of the distribution of posteriors $\overline{\mu}_1^*\equiv \int \mu F^*_1(d\mu)$ is uniformly absolutely continuous with respect to the prior $\mu_0^*$.
\end{theorem}
\noindent\emph{Proof of the ``if'' direction.} The proof of this direction is constructive. Given the measurable space $(X,\mathcal{X})$ and the true signal distribution $\mathbb{P}$, I construct the subjective distribution $\mathbb{Q}$ that rationalizes an observed pair $\big(\mu_0^*,F_1^*\big)$ that satisfies
\begin{equation}\label{eq:uniform_AC}
\overline{\mu}_1^*(D)\leq c\mu_0^*(D)
\end{equation}
for all $D\in\mathcal{X}$.\footnote{For any metric space $X$, the sigma-algebra on $X$ is assumed to be the Borel sigma-algebra, denoted by $\mathcal{X}$, and $\Delta(X)$ denotes the set of probability distributions on $(X,\mathcal{X})$, endowed with the topology of weak convergence and the corresponding Borel sigma-algebra.} Since $\overline{\mu}_1^*$ and $\mu_0^*$ are both probability measures on $(X,\mathcal{X})$, the constant $c$ in equation \eqref{eq:uniform_AC} must be at least weakly larger than one. If $c<1$, then $\overline{\mu}_1^*(X)<c\mu_0^*(X)=c<1$, a contradiction. If $c=1$, then $\overline{\mu}_1^*=\mu_0^*$. This is because $\overline{\mu}_1^*(D)\leq \mu_0^*(D)$ implies $\overline{\mu}_1^*(D^c)\geq \mu_0^*(D^c)$, where $D^c\in\mathcal{X}$ denotes the complement of $D$. But  $\overline{\mu}_1^*(D^c)\leq \mu_0^*(D^c)$ by equation \eqref{eq:uniform_AC}. Therefore, $\overline{\mu}_1^*(D)=\mu_0^*(D)$. Since $D$ is an arbitrary measurable set, $\overline{\mu}_1^*=\mu_0^*$. In the remainder of the proof, I construct the subjective distribution $\mathbb{Q}$ that rationalizes $(\mu_0^*,F_1^*)$, separately for the $c>1$ and $c=1$ cases.

I first prove the result for the $c>1$ case. I start by constructing the regular conditional probability $\nu:S\times\mathcal{X}\to [0,1]$ that represents agents' posterior about the state $x\in X$ conditional on the signal $s$. Let $\ominus\in S$ denote a signal such that $\mathbb{P}(\ominus)=0$. Such a signal always exists since $S=\Delta(X)$ is uncountable, but there are at most countably many signals $s\in S$ such that $\mathbb{P}(s)>0$. For any $s\in S$ such that $\varphi(s)\in\supp F^*_{1}$ and $s\neq \ominus$, set $\nu(s,D)=\varphi(s)(D)$ for all $D\in\mathcal{X}$. Set 
\[
\nu(\ominus,D) = \frac{c}{c-1}\mu_{0}^*(D)-\frac{1}{c-1}\overline{\mu}_1^*(D)
\]
for all $D\in\mathcal{X}$.
Finally, set $\nu(s,D)=\mu_0^*(D)$ for any $s\in S$ such that $\varphi(s)\notin \supp F^*_{1}\cup\{\ominus\}$ and all $D\in\mathcal{X}$, indicating that agents' posterior equals their prior conditional on any signal realized with zero probability. Note that, by construction, the mapping $s\mapsto\nu(s,D)$ is measurable for any $D\in\mathcal{X}$. Therefore, to show that $\nu$ is a kernel, it is sufficient to show that $\nu(s,\cdot)$ is a probability distribution on $(X,\mathcal{X})$ for all $s\in S$. This holds by construction for all $s\neq \ominus$. In order to verify that $\nu(\ominus,\cdot)$ is a probability measure, first note that, 
\begin{align*}
\nu\big(\ominus,X\big)& =\frac{c}{c-1}\mu_{0}^*(X)-\frac{1}{c-1}\overline{\mu}_1^*(X)=1,
\end{align*}
and 
\begin{align*}
\nu\big(\ominus,\emptyset\big) & =\frac{c}{c-1}\mu_{0}^*(\emptyset)-\frac{1}{c-1}\overline{\mu}_1^*(\emptyset)=0,
\end{align*}
where I am using the facts that $\mu_{0}^*(X)=\overline{\mu}_1^*(X)=1$ and $\mu_{0}^*(\emptyset)=\overline{\mu}_1^*(\emptyset)=0$ since $\mu_0^*$ and $\overline{\mu}^*_1$ are both probability measures on $X$. Next note that, for any set $D\in \mathcal{X}$,
\begin{align*}
\nu\big(\ominus,D\big) & = \frac{c}{c-1}\mu_{0}^*(D)-\frac{1}{c-1}\overline{\mu}_1^*(D)\geq \frac{c}{c-1}\mu_{0}^*(D)-\frac{c}{c-1}\mu_0^*(D)=0,
\end{align*}
where the inequality follows equation \eqref{eq:uniform_AC}. Finally, $\nu(\ominus,\cdot)$ is countably additive since both $\mu_{0}^*$ and $\overline{\mu}_1^*$ are probability measures and thus are countably additive. Therefore, $\nu(\ominus,\cdot)$ is a probability measure on $(X,\mathcal{X})$. 

I can now define the subjective distribution $\mathbb{Q}$, starting with its $S$-marginal distribution $\mathbb{Q}_S$. Let 
\[
\mathbb{Q}_S(E)\equiv\frac{1}{c}\mathbb{P}(E)+\frac{c-1}{c}\mathbb{1}\{\ominus\in E\},
\]
for all $E\in\mathcal{S}$, and let
\begin{equation}\label{eq:Bayes_rule_proof}
\mathbb{Q}(D\times E) \equiv \int_E \nu(s,D)\mathbb{Q}_S(ds)
\end{equation}
for all $D\in\mathcal{X}$ and $E\in \mathcal{S}$. Note that since the sigma-algebra $(\mathcal{X}\times\mathcal{S})$ over $(X\times S)$ is generated by sets of the form $D\times E$ with $D\in\mathcal{X}$ and $E\in \mathcal{S}$, the above expression fully specifies the probability distribution $\mathbb{Q}$. Moreover, comparing equations \eqref{eq:Bayes_rule} and \eqref{eq:Bayes_rule_proof} shows that $\nu$ is indeed a regular conditional probability of $\mathbb{Q}$ given $\mathcal{S}$. Lastly, since $\mathbb{Q}_S(E)=\frac{1}{c}\mathbb{P}(E)+\frac{c-1}{c}\mathbb{1}\{\ominus\in E\}\geq \frac{1}{c}\mathbb{P}(E)$ for all $E\in\mathcal{S}$, the true distribution $\mathbb{P}$ is uniformly absolutely continuous with respect to $\mathbb{Q}_S$.

It remains to show that $\mathbb{Q}_X=\mu_0^*$ and that the distribution of posteriors $F_\nu$, defined in equation \eqref{eq:population_dist}, coincides with the observed posterior distribution $F_1^*$. Note that, by definition, $F_1^*=\mathbb{P}\circ\varphi^{-1}$. Let 
\[
\hat{S}\equiv \supp \mathbb{P}= \{s\in S:\varphi(s)\in\supp F_1^*\}.
\] 
For any $D\in\mathcal{X}$,
\begin{align*}
   \mathbb{Q}_X(D)& =\int_S \nu(s,D)\mathbb{Q}_S(ds)\\ & = \frac{1}{c}\int_{\hat{S}}\varphi(s)(D)\mathbb{P}(ds)+\frac{c-1}{c}\nu(\ominus,D)\\ & = \frac{1}{c}\int_{\varphi(\hat{S})}\mu(D)\mathbb{P}\circ\varphi^{-1}(d\mu)+\mu_0^*(D)-\frac{1}{c}\overline{\mu}_1^*(D)\\ & = \frac{1}{c}\int_{\supp F_1^*}\mu(D) F_1^*(d\mu)+\mu_0^*(D)-\frac{1}{c}\overline{\mu}_1^*(D)\\ & = \frac{1}{c}\overline{\mu}^*_1(D)+\mu_0^*(D)-\frac{1}{c}\overline{\mu}^*_1(D)=\mu_0^*(D).
\end{align*}
I next show that $F_\nu=F_1^*$. Note that
\[
\mathbb{P}\left(\left\{s\in S: \varphi(s)\notin \supp F_1^*\right\}\right)=\mathbb{P}\left(\left\{s\in S: s\notin \supp \mathbb{P}\right\}\right)=0.
\]
On the other hand, by construction, $\mathbb{P}(\ominus)=0$. Therefore, for any $E\in\mathcal{S}$,
\begin{align*}
F_\nu(E) & =\mathbb{P}\left(\{s\in S:\nu(s,\cdot)\in E\}\right)\\ & =\mathbb{P}\left(\{s\in S:\nu(s,\cdot)\in E,\varphi(s)\in\supp F_1^*,s\neq \ominus\}\right)\\ & =\mathbb{P}\left(\{s\in S:\varphi(s)\in E\}\right)=F_1^*(E).
\end{align*}
This completes the proof for the $c>1$ case.

I next consider the case where $c=1$. I construct the regular conditional probability $\nu:S\times\mathcal{X}\to [0,1]$ by setting $\nu(s,D)=\varphi(s)(D)$ for all $s\in S$ such that $\varphi(s)\in\supp F^*_{1}$ and all $D\in\mathcal{X}$ and setting $\nu(s,D)=\mu_0^*(D)$ for all $s\in S$ such that $\varphi(s)\notin \supp F^*_{1}$ and all $D\in\mathcal{X}$. By construction, $\nu(s,\cdot)$ is a probability distribution on $(X,\mathcal{X})$, and the mapping $s\mapsto \nu(s,D)$ is measurable for all $D\in\mathcal{X}$. I set the $S$-marginal $\mathbb{Q}_S$ of the subjective distribution equal to the true distribution $\mathbb{P}$ of signals and define $\mathbb{Q}$ as in \eqref{eq:Bayes_rule_proof}. By construction, $\nu$ is a regular conditional probability of $\mathbb{Q}$ given $\mathcal{S}$, and $\mathbb{P}$ is uniformly absolutely continuous with respect to $\mathbb{Q}_S$. Next, note that, for any $D\in\mathcal{X}$,
\[
\mathbb{Q}_X(D) = \int_S \nu(s,D)\mathbb{P}(ds) = \int_{\supp F_1^*} \mu(D)F_1^*(d\mu)=\overline{\mu}_1^*(D)=\mu_0^*(D),
\]
where the last equality follows the fact that $\overline{\mu}_1=\mu_0^*$ when $c=1$, established in the first paragraph of the proof. Moreover, by an argument similar to the $c>1$ case, 
\[
F_\nu(E) = \mathbb{P}\left(\{s\in S:\nu(s,\cdot)\in E\}\right) =\mathbb{P}\left(\{s\in S:\varphi(s)\in E\}\right)=F_1^*(E)
\]
for all $E\in\mathcal{S}$. This shows that the subjective distribution $\mathbb{Q}$ constructed above rationalizes the observed pair $(\mu_0^*,F_1^*)$.\hfill\qed

\vspace{1em}
\noindent\emph{Proof of the ``only if'' direction.} Let $\mathbb{Q}$ denote agents' subjective distribution on $(X\times S)$ and $\nu$ denote the regular conditional probability of $\mathbb{Q}$ given $\mathcal{S}$. The existence of $\nu$ follows the assumption that $\mathbb{Q}$ satisfies condition (c) of Definition \ref{def:Bayesianism}. Since $\mathbb{Q}$ satisfies condition (a) and $\nu$ is a regular conditional probability of $\mathbb{Q}$ given $\mathcal{S}$,
\[
\mu_0^*(D) = \mathbb{Q}_X(D) = \int_{S}\nu(s,D)\mathbb{Q}_S(ds)
\]
for all $D\in\mathcal{X}$. On the other hand, for any $D\in\mathcal{X}$,
\[
\overline{\mu}_1^*(D) = \int_S \mu(D)F_1^*(d\mu)=\int_S \mu(D)F_\nu(d\mu)=\int_S\nu(s,D)\mathbb{P}(ds),
\]
where the second equality is due to the assumption that $\mathbb{Q}$ satisfies condition (c) of Definition \ref{def:Bayesianism}, and the third equality follows the definition of $F_\nu$. 
Finally, $\mathbb{P}$ is uniformly absolutely continuous with respect to $\mathbb{Q}_S$ by condition (b) of Definition \ref{def:Bayesianism}. Therefore, by Lemma \ref{lem:unif_abs_cont} of the appendix, there exists a Radon--Nikodym derivative $f\equiv \frac{d\mathbb{P}}{d\mathbb{Q}_S}$ that is bounded up to sets of $\mathbb{Q}_S$-measure zero. Therefore,
\[
\int_{S}\nu(s,D)\mathbb{P}(ds)=\int_{S}\nu(s,D)f(s)\mathbb{Q}_S(ds) \leq c \int_{S}\nu(s,D)\mathbb{Q}_S(ds),
\]
where $c$ denotes a positive constant such that $f\leq c$ with $\mathbb{Q}_S$-probability one. Combining the last three displays implies that $\overline{\mu}^*_1(D)\leq c\mu_0^*(D)$. The fact that $c$ can be chosen independently of the set $D\in\mathcal{X}$ establishes that $\overline{\mu}^*_1$ is uniformly absolutely continuous with respect to $\mu_0^*(D)$ and completes the proof.\hfill\qed

The uniform absolute continuity of the posterior mean with respect to the prior encompasses the entire empirical content of Bayesianism. Absent additional a priori restrictions on what constitutes a reasonable subjective distribution, any belief sequence that satisfies this condition is consistent with Bayesian updating. It is easy to see that the absolute continuity condition is necessary for Bayesianism: If the prior of a Bayesian agent assigns zero probability to an event, her posterior must also assign zero probability to the event---regardless of agents' subjective belief and the true distribution of signals. What is more surprising is that absolute continuity is also sufficient for consistency with Bayesianism. The proof establishes this result by constructing a subjective distribution $\mathbb{Q}$ starting from a pair $(\mu_0^*,F_1^*)$, which satisfies the absolute continuity condition, and showing that the conjectured $\mathbb{Q}$ satisfies the properties set out in Definition \ref{def:Bayesianism}.

\subsection{Special Cases}
Building on Theorem \ref{thm:main}, I next discuss two of its corollaries. The first corollary addresses scenarios where the state space $X$ is finite, a common situation in applications. The corollary follows from the theorem given the following observation: For two distributions $P$ and $Q$ over a finite set, $P$ is uniformly absolutely continuous with respect to $Q$ if the support of $P$ is a subset of the support of $Q$.
\begin{corollary}\label{cor:finite_support}
Suppose the state space $X$ is finite. A pair $(\mu_0^*,F_1^*)$ is consistent with Bayesianism if and only if 
\[
\supp\overline{\mu}_1^*\subseteq \supp\mu_0^*,
\]
where $\overline{\mu}_1^*\equiv \int \mu F^*_1(d\mu)$ denotes the mean of the distribution of posteriors.
\end{corollary}
This result summarizes the empirical content of Bayesianism in discrete settings. The mean posterior belief cannot assign positive probabilities to states that have zero probability according to the prior. The necessity of this property for beliefs to be consistent with Bayesianism is apparent given Bayes' rule. The corollary goes a step further by establishing that this property is also sufficient for belief sequences to be consistent with Bayesianism. A further corollary of Corollary \ref{cor:finite_support} is that \emph{any} posterior distribution is consistent with Bayesianism when the prior has full support over a finite state space.
\begin{corollary}\label{cor:finite_full_support}
Suppose the state space $X$ is finite and $\mu_0^*$ has full support over $X$. Then the pair $(\mu_0^*,F_1^*)$ is consistent with Bayesianism for any distribution $F_1^*$ of posteriors.
\end{corollary}

The following example illustrates an application of the theorem and its corollaries:
\begin{example}\label{example1}
The state takes values in the set $X=\{H,L\}$. Agents' observed prior is the uniform distribution over $X$. The observed distribution of posteriors $F_1^*$ is as follows: For a quarter of agents, the belief that the state is $H$ goes up to $0.8$. The remaining agents become certain that the state is $H$. Should the observation of this belief sequence lead the analyst to conclude that agents are not Bayesian? The answer is no; the observed pair $(\mu_0^*,F_1^*)$ is indeed consistent with Bayesianism. This conclusion follows Corollary \ref{cor:finite_full_support} of Theorem \ref{thm:main} by noting that $\mu_0^*$ has full support over $X$. 

The observation of $F_1^*$ imposes some restrictions on the true distribution of signals $\mathbb{P}$ and the mapping $\varphi$ used by agents to update their beliefs. Since agents' posteriors take on two values, there are at least two signals that are realized with positive probabilities. The observation of one set of signals moves agents' posterior belief that the state is $H$ to $0.8$. Since a quarter of agents end up with the posterior $\mu_1(H)=0.8$, the signals that lead to this posterior must have probability $\mathbb{P}(\{s:\varphi(s)=(\mu_1(H)=0.8)\})=0.25$. Likewise, there is a set of signals that has true probability $\mathbb{P}(\{s:\varphi(s)=(\mu_1(H)=1)\})=0.75$ and makes agents certain that the state is $H$. With slight abuse of notation, in the remainder of the example, I refer to the $\{s:\varphi(s)=(\mu_1(H)=0.8)\}$ and $\{s:\varphi(s)=(\mu_1(H)=1)\}$ events simply as the $s=0.8$ and $s=1$ signals, respectively.\footnote{This would not be an abuse of notation under the assumption that $\varphi$ is the identity mapping. Note that the assumption that $\varphi$ is the identity mapping is innocuous in this example since observing $F_1^*$ only identifies $\mathbb{P}\circ \varphi^{-1}=F_1^*$---but not $\mathbb{P}$ or 
$\varphi$. Nonetheless, the construction in the example can be easily modified to allow for the possibility that $F_1^*$, $\mathbb{P}$, and $\varphi$ are separately identified by the analyst. I forgo this extension here since it would lead to additional notational complexity without offering any new insights. See the proof of Theorem \ref{thm:main} for the general construction.}

I illustrate how $(\mu_0^*,F_1^*)$ can be rationalized by finding a subjective distribution $\mathbb{Q}$ such that the belief sequence of a Bayesian agent with subjective distribution $\mathbb{Q}$ matches the observed prior and distribution of posteriors. The distribution $\mathbb{Q}$ needs to satisfy three requirements for it to rationalize the observed prior $\mu_0^*$ and posterior distribution $F_1^*$. First, $\mathbb{Q}$ must be consistent with the observed prior $\mu_0^*$; i.e., $\mathbb{Q}(H)=\mu_0^*(H)=0.5$.  Second, agents must assign positive probabilities to the $s=0.8$ and $s=1$ signals for Bayes' rule to be applicable after the observation of those signals. Third, agents' posterior conditional on the observation of  the $s=0.8$ and $s=1$ signals must be consistent with their observed posteriors; i.e., $\mathbb{Q}(H|s=0.8)=0.8$ and $\mathbb{Q}(H|s=1.0)=1.0$. 

One needs to also specify what agents believe about the probability of observing signals other than $0.8$ and $1.0$. I start by assuming that agents believe the signal can only take values $s=0.8$ and $s=1.0$. This assumption constrains $\mathbb{Q}$ to satisfy $\mathbb{Q}(\{(x,s):s\in \{0.8,1.0\}\})=1$. This constraint, together with the requirements previously discussed and the requirement that $\mathbb{Q}(x,s)\geq 0$ for any $(x,s)$, yields a mixed system of equalities and inequalities for the four unknown probabilities $\mathbb{Q}(H,0.8)$, $\mathbb{Q}(L,0.8)$, $\mathbb{Q}(H,1.0)$, and $\mathbb{Q}(L,1.0)$:
\begin{align}
    & \mathbb{Q}(H,0.8)+\mathbb{Q}(L,0.8)>0,\label{eq:eg_requirment_1}\\
    & \mathbb{Q}(H,1.0)+\mathbb{Q}(L,1.0)>0,\\
    & \frac{\mathbb{Q}(H,0.8)}{\mathbb{Q}(H,0.8)+\mathbb{Q}(L,0.8)}=0.8,\\
    & \frac{\mathbb{Q}(H,1.0)}{\mathbb{Q}(H,1.0)+\mathbb{Q}(L,1.0)}=1.0,\label{eq:eg_requirment_4}\\
    &  \mathbb{Q}(H,0.8), \mathbb{Q}(L,0.8), \mathbb{Q}(H,1.0), \mathbb{Q}(L,1.0) \geq 0,\label{eq:eg_requirment_7}\\
    & \mathbb{Q}(H,0.8)+\mathbb{Q}(H,1.0)=0.5,\label{eq:eg_sum_to_half_H}\\
    & \mathbb{Q}(L,0.8)+\mathbb{Q}(L,1.0)=0.5.\label{eq:eg_sum_to_half_L}
\end{align}
It is easy to verify that this system does not have a solution. 

Thus, for the observed belief sequence to arise from Bayesian updating by agents, they must believe the possibility that the signal takes values outside the set $\{0.8,1.0\}$. With slight abuse of notation, I let $s=\ominus$ denote the event that the signal takes a value outside the set $\{0.8,1.0\}$. Constraints \eqref{eq:eg_sum_to_half_H} and \eqref{eq:eg_sum_to_half_L} now have to be modified as follows:
\begin{align}
    & \mathbb{Q}(H,0.8)+\mathbb{Q}(H,1.0)+\mathbb{Q}(H,\ominus)=0.5,\label{eq:eg_sum_to_half_H_prime}\\
    & \mathbb{Q}(L,0.8)+\mathbb{Q}(L,1.0)+\mathbb{Q}(H,\ominus)=0.5.\label{eq:eg_sum_to_half_L_prime}
\end{align}
The remaining requirements, expressed in equations  \eqref{eq:eg_requirment_1}--\eqref{eq:eg_requirment_7}, remain intact. However, $\mathbb{Q}$ must now additionally satisfy the two non-negativity requirements:
\begin{align}
    & \mathbb{Q}(H,\ominus),\mathbb{Q}(L,\ominus)\geq 0.\label{eq:eg_requirment_5}
\end{align}
Equations \eqref{eq:eg_requirment_1}--\eqref{eq:eg_requirment_7} and \eqref{eq:eg_sum_to_half_H_prime}--\eqref{eq:eg_requirment_5} constitute a mixed system of equalities and inequalities for the six unknown probabilities $\mathbb{Q}(H,0.8)$, $\mathbb{Q}(L,0.8)$, $\mathbb{Q}(H,1.0)$, $\mathbb{Q}(L,1.0)$, $\mathbb{Q}(H,\ominus)$, and $\mathbb{Q}(L,\ominus)$. The fact that the mean of the posterior distribution has the same support as the prior is sufficient to ensure that this system has a solution. One such solution---and the one corresponding to the proof of Theorem \ref{thm:main}---is as follows:
\settowidth{\mycolwd}{$0.1875$}
\[
\renewcommand\arraystretch{2}
\begin{array}{l|C{\mycolwd}|C{\mycolwd}|C{\mycolwd}|}
\multicolumn{1}{r}{}
 & \multicolumn{1}{c}{0.8}
 & \multicolumn{1}{c}{1.0}
 & \multicolumn{1}{c}{\ominus}\\
\cline{2-4}
H & \hfill0.25\hfill\hfill  & \hfill0.25\hfill\hfill & \hfill0\hfill\hfill\\
\cline{2-4}
L & 0.0625 & \hfill 0 \hfill\hfill & \hfill 0.4375\hfill\\
\cline{2-4}
\end{array}
\]

Note that the observed pair $(\mu_0^*,F_1^*)$ can be rationalized only if agents are allowed to have a misspecified belief about the distribution of signals. If agents were to hold a correctly specified belief, the subjective distribution $\mathbb{Q}$ would have to agree with the true distribution $\mathbb{P}$ on the probabilities of different signals. But then the systems of equalities and inequalities that determine $\mathbb{Q}$ would have no solution.
\end{example}

\section{Discussion}
This section discusses the assumptions that underpin the paper's main result and how relaxing them changes the conclusion.

\subsection{Assumptions}\label{sec:assumptions_discussion}
The paper makes a number of explicit and implicit assumptions in order to obtain the characterization in Theorem~\ref{thm:main}. These assumptions can be divided into three categories:
\begin{itemize}
    \item[1.] Homogeneity assumptions: Agents are ex ante identical, use the same function to map their signals to posteriors, and receive i.i.d. signals.
    \item[2.] Knowledge assumptions: The analyst directly and perfectly observes agents' beliefs---instead of having to identify them from choice data. He knows the true distribution of signals and the mapping agents use to update their beliefs. He observes the population distribution of posteriors.
    \item[3.] No restrictions on the subjective belief: The analyst puts no restrictions on what constitutes a reasonable subjective belief. Moreover, the analyst does not observe agents' beliefs concerning the likelihood of various signals.
\end{itemize}

These sets of assumptions serve different purposes. The homogeneity assumptions help with the identification of agents' subjective beliefs. The analyst observes each agent only after the realization of a single signal. Without the homogeneity assumptions, how an agent behaves after a signal would not be informative of how other agents would have behaved if they observed that same signal.

The knowledge assumptions limit what the analyst can freely choose in order to rationalize his observations. When the analyst knows an object such as the true signal distribution, he has no flexibility in choosing that object to rationalize agents' behavior. This limits the set of observations that can be rationalized by the analyst and makes it easier for him to reject his observations as consistent with Bayesianism.

The first two sets of assumptions strengthen the ``if'' direction of Theorem \ref{thm:main}. Under these assumptions, the analyst cannot reject agents' Bayesianism as long as his observations satisfy the uniform absolute continuity condition. This remains true despite the constraints these assumptions place on the analyst's ability to rationalize observations. If these assumptions were to be relaxed, certain observations---even those that do not adhere to the uniform absolute continuity condition---might still be explainable as consistent with Bayesianism. Hence, the uniform absolute continuity condition may be seen as an upper bound on the empirical content of Bayesianism. The main role of the first two sets of assumptions is to make this bound tight by turning the theorem into an ``if and only if'' result. 

The final set of assumptions enables the analyst to freely choose the subjective belief $\mathbb{Q}$ to rationalize his observations. The analyst is assumed to perfectly observe what agents believe about the state but not how they intend to update those beliefs. Therefore, he is free to form any conjecture about how agents are planning to do so. This degree of freedom makes it more challenging for the analyst to disprove the Bayesianism of agents. Unlike the first two sets of assumptions, which narrow the range of observations consistent with Bayesianism, the third set introduces more flexibility, thus expanding this range.

The assumption that any well-defined subjective distribution is permissible is made in keeping with \citet*{savage1972foundations}'s idea of purely subjective beliefs. I assume that any subjective belief $\mathbb{Q}$ that is consistent with agents' elicited beliefs is a  valid subjective distribution. Agents' rationality is judged not based on their subjective beliefs but based on how they update those beliefs. In the next subsection, I discuss two alternatives to this assumption proposed in the literature and how they change the conclusion of Theorem \ref{thm:main}.

\subsection{Alternative Notions of Bayesianism}
The first alternative I consider assumes that agents have a correctly specified belief about the distribution of signals. This assumption leads to the martingale property of Bayesian beliefs: The expectation of the posterior must equal the prior. \citet*{aumann1995repeated} and \citet*{Kamenica2009} show that this is indeed the only restriction imposed on beliefs by the requirement that agents are Bayesian. The following theorem generalizes this result to general metric spaces. More importantly, however, it highlights the fact that the martingale property characterizes the empirical content of Bayesianism \emph{only} when agents are required to have correct beliefs about the distribution of signals.
\begin{theorem}
The pair $(\mu_0^*,F_1^*)$ is consistent with Bayesianism given a subjective distribution $\mathbb{Q}$ with the $S$-marginal satisfying $\mathbb{Q}_S=\mathbb{P}$ if and only if $\mu_0=\overline{\mu}_1^*\equiv \int \mu F^*_1(d\mu)$.
\end{theorem}
\begin{proof} Proof of the ``if'' direction is identical to the $c=1$ case in the proof of Theorem \ref{thm:main}'s ``if'' direction. The proof of the ``only if'' direction follows similar steps as in the proof of Theorem \ref{thm:main}'s ``only if'' direction. In particular, by an identical argument, $\mu_0^*(D) =\int_{S}\nu(s,D)\mathbb{Q}_S(ds)$ and $\overline{\mu}_1^*(D) =\int_S\nu(s,D)\mathbb{P}(ds)$. The assumption that $\mathbb{Q}_S=\mathbb{P}$ completes the proof.
\end{proof}

A more permissive notion of Bayesianism is proposed by \citet*{Shmaya2016}. They allow Bayesian agents to have incorrect beliefs about the distribution of signals--- as long as the supports of those beliefs coincide with the support of the true distribution. The following theorem generalizes \citet*{Shmaya2016}'s Lemma 1 to general metric state spaces and arbitrary true signal distributions. It reduces to their result when both $X$ and $\supp\mathbb{P}$ are finite sets. However, its main significance is to clarify that \citet*{Shmaya2016}'s conclusion relies on an a priori restriction on what constitutes a reasonable subjective distribution.

\begin{theorem}
The following statements are equivalent:
\begin{itemize}
    \item[A.] The pair $(\mu_0^*,F_1^*)$ is consistent with Bayesianism given a subjective distribution $\mathbb{Q}$ with a $S$-marginal $\mathbb{Q}_S$ that is uniformly absolutely continuous with respect to $\mathbb{P}$.
    \item[B.] There exists a probability measure $\lambda \in \Delta(S)$ such that $\lambda$ and $F_1^*$ are mutually uniformly absolutely continuous and $\mu_0^*=\int \mu\lambda(d\mu)$.
\end{itemize}
\end{theorem}

\begin{proof}
First, suppose there exists a probability measure $\lambda \in \Delta(S)$ such that $\lambda$ and $F_1^*$ are mutually uniformly absolutely continuous and $\mu_0^*=\int \mu\lambda(d\mu)$. By Lemma \ref{lem:unif_abs_cont} of the appendix, there exist Radon--Nikodym derivatives $f\equiv \frac{d\lambda}{dF_1^*}$ and $\frac{1}{f}\equiv \frac{dF_1^*}{d\lambda}$ such that $c\leq f\leq C$ for some positive constants $c,C$ (up to sets of $F_1^*$- and $\lambda$-measure zero). Set $\nu(s,D)=\varphi(s)(D)$ for all $s\in S$ and $D\in\mathcal{X}$, and set $\mathbb{Q}_S(ds)=f(\varphi(s))\mathbb{P}(ds)$. I need to show that $\mathbb{Q}_S$, as defined above, is indeed a probability distribution on $(S,\mathcal{S})$. By construction, $\mathbb{Q}_S(E)\geq 0$ for all $E\in\mathcal{S}$, and $\mathbb{Q}_S(\emptyset)=0$. Next, note that
\[
\int_S \mathbb{Q}_S(ds) = \int_S f(\varphi(s))\mathbb{P}(ds)=\int_S f(\mu) \mathbb{P}\circ \varphi^{-1}(d\mu)=\int_Sf(u)F_1^*(d\mu)=\int_S \lambda(d\mu)=1,
\]
where the first equality is by definition, the second one uses the change-of-variables formula for pushforward measures, the third equality is due to the fact that $F_1^*=\mathbb{P}\circ\varphi^{-1}$, the fourth one uses the definition of $f$, and the last equality is because $\lambda$ is a probability measure on $S$. Finally, $\mathbb{Q}_S$ is countably additive since $\mathbb{P}$ is countably additive. Therefore, $\mathbb{Q}_S$ is a well-defined probability distribution. I finish the construction by defining $\mathbb{Q}$ as in equation \eqref{eq:Bayes_rule}. Note that, by construction, $\nu$ is a conditional probability of $\mathbb{Q}$ given $\mathcal{S}$. Furthermore, by an argument similar to the one in the above display,
\begin{align*}
\mathbb{Q}_X & =\int_S \nu(s,\cdot)\mathbb{Q}_S(ds)=\int_S\varphi(s)f(\varphi(s))\mathbb{P}(ds)= \int_S \mu f(\mu)F_1^*(d\mu) = \int_S \mu \lambda(\mu)=\mu_0^*,
\end{align*}
where the last equality is by assumption. Therefore, condition (a) of Definition \ref{def:Bayesianism} is satisfied. Furthermore, since $\mathbb{Q}_S(ds)=f(\varphi(s))\mathbb{P}(ds)$ and $c\leq f\leq C$ almost surely, $\mathbb{Q}_S$ and $\mathbb{P}$ are mutually uniformly absolutely continuous. That is, condition (b) of Definition \ref{def:Bayesianism} is satisfied, and $\mathbb{Q}_S$ is uniformly absolutely continuous with respect to $\mathbb{P}$. On the other hand,
\[
F_\nu(E) = \mathbb{P}\left(\{s\in S:\nu(s,\cdot)\in E\}\right) =\mathbb{P}\left(\{s\in S:\varphi(s)\in E\}\right)=\mathbb{P}\circ\varphi^{-1}(E)=F_1^*(E)
\]
for all $E\in\mathcal{S}$, implying that condition (c) is also satisfied.

Next, suppose $(\mu_0^*,F_1^*)$ is consistent with Bayesianism given a subjective distribution $\mathbb{Q}$ with a $S$-marginal $\mathbb{Q}_S$ that is uniformly absolutely continuous with respect to $\mathbb{P}$, and let $\nu$ denote the regular conditional probability of $\mathbb{Q}$ given $\mathcal{S}$. The existence of $\nu$ follows the assumption that $\mathbb{Q}$ satisfies condition (b) of Definition \ref{def:Bayesianism}. I define $\lambda\in\Delta(S)$ as follows:
\[
\lambda(E) \equiv  \mathbb{Q}_S(\{s\in S: \nu(s,\cdot)\in E\})
\]
for all $E\in\mathcal{S}$. I next show that $\lambda$ and $F_1^*$ are mutually uniformly absolutely continuous and $\mu_0^*=\int \mu\lambda(d\mu)$. Since $\mathbb{Q}$ satisfies condition (a) and $\nu$ is a regular conditional probability of $\mathbb{Q}$ given $\mathcal{S}$,
\[
\mu_0^* = \mathbb{Q}_X = \int_{S}\nu(s,\cdot)\mathbb{Q}_S(ds)=\int_S \mu\lambda(d\mu),
\]
where the last equality is by definition. On the other hand, for all $E\in\mathcal{S}$,
\[
\lambda(E)=\mathbb{Q}_S(\{s\in S: \nu(s,\cdot)\in E\})\geq \frac{1}{c}\mathbb{P}(\{s\in S: \nu(s,\cdot)\in E\})=\frac{1}{c}F_\nu(E)=\frac{1}{c}F_1^*(E)
\]
for some positive constant $c$, where the first two equalities are by definition, the inequality is by condition (b) of Definition \ref{def:Bayesianism}, and the third equality is by condition (c) of Definition \ref{def:Bayesianism}. Likewise, since $\mathbb{Q}_S$ is uniformly absolutely continuous with respect $\mathbb{P}$, for all $E\in\mathcal{S}$,
\[
\lambda(E)=\mathbb{Q}_S(\{s\in S: \nu(s,\cdot)\in E\})\leq C\mathbb{P}(\{s\in S: \nu(s,\cdot)\in E\})=CF_\nu(E)=CF_1^*(E)
\]
for some positive constant $C$. The fact that $c$ and $C$ can be chosen independently of $E\in\mathcal{S}$ in the above two displays establishes that $\lambda$ and $F_1^*$ are mutually uniformly absolutely continuous.
\end{proof}

\appendix

\section*{Technical Appendix}
\begin{lemma}\label{lem:unif_abs_cont}
Let $P$ and $Q$ be probability distributions over the same measurable space. $P$ is uniformly absolutely continuous with respect to $Q$ if and only if there exists a Radon--Nikodym derivative $f\equiv \frac{dP}{dQ}$ and a positive constant $c$ such that $f\leq c$ up to sets of $Q$-measure zero.
\end{lemma}

\begin{proof}
First, suppose there exists a Radon--Nikodym derivative $f\equiv \frac{dP}{dQ}$ and a positive constant $c$ such that $f\leq c$ up to sets of $Q$-measure zero. For any measurable set $E$,
\[
P(E) = \int_E dP = \int_E f dQ \leq c \int_E dQ = cQ(E).
\]
That is, $P$ is uniformly absolutely continuous with respect to $Q$.

Next, suppose $P$ is uniformly absolutely continuous with respect to $Q$. Then, by definition, $P(E)\leq cQ(E)$ for some $c$ and any measurable set $E$. In particular, $P(E)=0$ for any $E$ for which $Q(E)=0$. Therefore, $P$ is absolutely continuous with respect to $Q$, and so, by the Radon--Nikodym theorem, there exists a derivative $f\equiv \frac{dP}{dQ}$. I finish the proof by showing that $f$ is bounded $Q$-almost surely. Toward a contradiction, suppose that for any positive constant $C$ there exists a measurable set $E$ with $Q(E)>0$ such that $f>C$ on $E$. Then,
\[
P(E)=\int_E dP = \int_E f dQ>C\int_E dQ=CQ(E).
\]
Since $C$ is arbitrary, there exists no constant $c$ such that $P(E)\leq cQ(E)$ for all $E$, a contradiction to the assumption that $P$ is uniformly absolutely continuous with respect to $Q$.
\end{proof}

\newpage

\fontsize{11}{11}\selectfont\baselineskip0.52cm
\bibliographystyle{te}
\bibliography{subjective}

\end{document}